\documentclass[twocolumn,aps,prl,superscriptaddress,showpacs,showkeys]{revtex4}

\usepackage{CJK}%
\usepackage{color}%
\usepackage{amsmath,bm}%
\usepackage{graphicx}%
\begin{document}
\begin{CJK*}{GBK}{}

\title{Statistical analysis of experimental multifragmentation events in $^{64}$Zn + $^{112}$Sn at 40 MeV/nucleon}

\author{W. Lin}
\affiliation{Key Laboratory of Radiation Physics and Technology of the Ministry of Education, Institute of Nuclear Science and Technology, Sichuan University, Chengdu 610064,	China}
\affiliation{Institute of Modern Physics, Chinese Academy of Sciences, Lanzhou, 730000, China}
\author{H. Zheng}
\affiliation{Laboratori Nazionali del Sud, INFN, I-95123 Catania, Italy}
\author{P. Ren}
\affiliation{Key Laboratory of Radiation Physics and Technology of the Ministry of Education, Institute of Nuclear Science and Technology, Sichuan University, Chengdu 610064,	China}
\affiliation{Institute of Modern Physics, Chinese Academy of Sciences, Lanzhou, 730000, China}
\author{X. Liu}
\email[E-mail at:]{liuxingquan@impcas.ac.cn}
\affiliation{Institute of Modern Physics, Chinese Academy of Sciences, Lanzhou, 730000, China}
\author{M. Huang}
\affiliation{College of Physics and Electronics information, Inner Mongolia University for Nationalities, Tongliao, 028000, China}
\author{R. Wada}
\email[E-mail at:]{wada@comp.tamu.edu}
\affiliation{Cyclotron Institute, Texas A$\&$M University, College Station, Texas 77843}
\author{Z. Chen}
\affiliation{Institute of Modern Physics, Chinese Academy of Sciences, Lanzhou, 730000, China}
\author{J. Wang}
\affiliation{Institute of Modern Physics, Chinese Academy of Sciences, Lanzhou, 730000, China}
\author{G. Q. Xiao}
\affiliation{Institute of Modern Physics, Chinese Academy of Sciences, Lanzhou, 730000, China}
\author{G. Qu}
\affiliation{Key Laboratory of Radiation Physics and Technology of the Ministry of Education, Institute of Nuclear Science and Technology, Sichuan University, Chengdu 610064,	China}
\affiliation{Institute of Modern Physics, Chinese Academy of Sciences, Lanzhou, 730000, China}
\date{\today}

\begin{abstract}
  A statistical multifragmentation model (SMM) is
  applied to the experimentally observed multifragmentation events in an intermediate heavy ion reaction.
  Using the temperature and symmetry energy extracted from the isobaric yield ratio (IYR) method based on the Modified Fisher Model (MFM), SMM is applied to the reaction $^{64}$Zn + $^{112}$Sn at 40 MeV/nucleon. The experimental isotope distribution and mass distribution of the primary reconstructed fragments are compared without afterburner and they are
  well reproduced. The extracted temperature $T$ and symmetry energy coefficient $a_{sym}$ from SMM simulated events, using the IYR method, are also consistent with those from the experiment. These results strongly suggest that in the multifragmentation process there is a freezeout volume, in which the thermal and chemical equilibrium is established before or at the time of the intermediate-mass fragments emission.
\end{abstract}


\maketitle
\end{CJK*}

\section*{I. Introduction}
  In violent heavy ion collisions of central collisions in the intermediate energy regime (20 MeV/nucleon $< E_{inc} <$ a few hundred MeV/nucleon) and peripheral collisions in the relativistic energy regime (a few GeV/nucleon), intermediate-mass fragments (IMFs) are copiously produced in multifragmentation processes. Nuclear multifragmentation was predicted in 1930's ~\cite{Bohr36} and has been extensively studied following the advent of $4\pi$ detectors~\cite{Borderie08,Gulminelli06,Chomaz04,Scharenberg2001,Hongfei1997}.
  Nuclear multifragmentation occurs when a large amount of energy is deposited in a finite nucleus. In general, the nuclear multifragmentation process can be divided into three stages in intermediate heavy ion collisions, i.e., dynamical compression and expansion, the formation of primary hot fragments, and finally the separation and cooling of the primary hot fragments by statistical gamma and particle emissions.
  Nuclear multifragmentation is of great importance for the constraining of density dependence of symmetry energy, which plays a key role for various phenomena in nuclear	astrophysics, nuclear structure, and nuclear reactions~\cite{Lattimer04,BALi08,BotvinaNPA2010}. Moreover, the multifragmentation in relativistic heavy ion collisions also allows for producing a new kind of large nuclei - hypernuclei~\cite{BotvinaPRC2007,BotvinaPRC2017}.

  Different transport models have been developed to model the multifragmentation process. They are Boltzmann-Uehling-Uhlenbeck model (BUU)~\cite{Aichelin85}, the stochastic mean field model (SMF)~\cite{Colonna98,Baran12,Gagnon12}, Vlasov-Uehling-Uhlenbeck model (VUU)~\cite{Kruse85}, Boltzmann-Nordheim-Vlasov model (BNV)~\cite{Baran02}, quantum molecular dynamics model (QMD)~\cite{Peilert89,Aichelin91,Lukasik93}, constrained molecular dynamics model (CoMD)~\cite{Papa01,Papa05,Papa07,Papa09}, improved quantum molecular dynamics model (ImQMD)~\cite{Wang02,Wang04,Zhang05,Zhang06,Zhang12}, fermionic molecular dynamics model (FMD)~\cite{Feldmeier90}, antisymmetrized molecular dynamics model (AMD)~\cite{Ono96,Ono99,Ono02} among others. Most of these can account reasonably well for many characteristic properties of experimental observables.

  In transport models, simulated events for a given reaction system show large fluctuations in space and time for the formation of IMFs. The large fluctuation causes difficulty in identifying, on an event by event basis, a unique freezeout volume and time, when thermal and chemical equilibrium is established. However there are some evidences that statistical equilibrations are established before or at the time of the IMFs produced while the observables are averaged over many events. Furuta et al. demonstrated in Ref.~\cite{Furuta09} that, in AMD calculations of $^{40}$Ca + $^{40}$Ca at 35 MeV/nucleon, IMFs are formed in a wide range of time interval (100 fm/c $-$ 300 fm/c) and the isotope yield distributions change with time. However the yield and excitation energy distributions as a function of mass at a given time can be identified as one of statistically equilibrated ensembles generated by the same model separately. In Ref.~\cite{Ono03}, it is reported that isoscaling holds, which is not evident a priori in dynamical models, in a study of similar reaction systems.

  In our previous works, we presented that IMFs with mass $\ge$ 15 show a power law distribution with the critical exponent, $A^{-2.3}$, in the reconstructed primary fragments~\cite{Lin14,Lin14PRC89_021601R}. A self-consistent analysis for AMD events of $^{40}$Ca + $^{40}$Ca at energy range from 35 to 300 MeV/nucleon~\cite{Liu15} strongly suggests that the variety in the dynamical fragmentation process originates from the fluctuation of a statistical ensemble in time, that is, IMFs are formed at a freezeout volume, which is characterized by the time-- and event--averaged density and temperature. The density and temperature of an equilibrated ensemble are the basic assumptions of statistical multifragmentation models. The reconstructed primary fragment yields in $^{64}$Zn + $^{112}$Sn at 40 MeV/nucleon~\cite{Lin14,Lin14PRC89_021601R} also provide the direct comparisons with model calculations without secondary decay processes, which often cause complexity for the comparisons~\cite{Ono07,Fevre05}.

  In this paper we present a common feature of a freezeout concept between the dynamical transport model simulations and the statistical multifragmentation calculations, focusing on the formation of the IMFs in intermediate heavy ion collisions.

  The paper is organized as follows. In section II, the Modified Fisher model (MFM) and SMM with symmetry entropy are briefly described. The consistency between SMM with the symmetry entropy and the isobaric yield ratio (IYR) method based on MFM is presented in section III. Detail comparisons between SMM primary fragment yields and the experimental results of $^{64}$Zn + $^{112}$Sn at 40 MeV/nucleon are carried out in section IV. A summary is given in section V.

\section*{II. Models}

\section*{II.A MFM formulation}

  The Modified Fisher Model (MFM)~\cite{Fisher1967,Minich1982,Hirsch1984,Bonasera2008} is applied to characterize the emitting source of IMFs in the previous works~\cite{Liu14PRC90_014605,Lin14PRC89_021601R,Huang10PRC81_044620,Lin14,Liu15}. In the framework of MFM, the yield of an isotope with $I=N-Z$ and mass $A$ (N and Z are the numbers of neutrons and protons respectively) produced in a multifragmentation reaction, can be given as
  \begin{eqnarray}
  Y(I,A) =& Y_{0} A^{-\tau}\exp\left[\frac{W(I,A)+\mu_{n}N+\mu_{p}Z}{T}+S^{mix}_{A,Z}\right],
  \label{eq:eq_MFM}
  \end{eqnarray}
  where $S^{mix}_{A,Z}=-\ln(N!Z!/A!)\approx -[N\ln(N/A)+Z\ln(Z/A)]$ is the mixing entropy from the two components of nuclear matter at the classical limit of the non-interacting Fermi gas at the time of the fragment formation. $\mu_{n}$ ($\mu_{p}$) is the neutron (proton) chemical potential. Using the generalized Weizs\"{a}cker-Bethe semiclassical mass formula~\cite{Weizsacker1935,Bethe1936}, $W(I,A)$ can be approximated as
  \begin{eqnarray}
  \nonumber W(I,A) &=& a_{v}A- a_{s}A^{2/3}- a_{c}\frac{Z(Z-1)}{A^{1/3}}\\
  \nonumber & &-a_{sym}\frac{I^{2}}{A} - a_{p}\frac{\delta}{A^{1/2}},\\
  \delta &=& - \frac{(-1)^{Z}+(-1)^{N}}{2}.
  \label{eq:eq_WB}
  \end{eqnarray}
  In general the coefficients, $a_{v}$, $a_{s}$, $a_{sym}$, $a_{p}$ and the chemical potentials are temperature and density dependent, even though they are not shown explicitly.

  The isobaric yield ratio method, based on the MFM, proposed in Ref.~\cite{Huang10PRC81_044620} allows one to extract $a_{sym}/T$ from the yield ratio of two pairs of isobars produced in the same reaction system, $R(I+2,I,A) = Y(I+2,A)/Y(I,A)$, as
  \begin{eqnarray}\label{eq:asym_T}
  \nonumber \frac{a_{sym}}{T} &=& -\frac{A}{8}\left\{\ln[R(3,1,A)]-\ln[R(1,-1,A)]\right.\\
  & &\left.-\Delta(3,1,A)+\Delta E_{c}\right\}.
  \end{eqnarray}
  $\Delta(3,1,A)$ is the difference in the mixing entropy of isobars A with $I$ = 3 and 1; $\Delta E_{c} = 2a_{c}/(A^{1/3}T)$ is the difference of Coulomb energy between neighboring isobars. The Coulomb energy coefficient relative to temperature and the yield ratio of isobar A with $I$ = 1 and $-1$ are related by the following equation as
  \begin{eqnarray}\label{eq:ac_T}
  \ln[R(1,-1,A)] &=& [\Delta\mu + 2a_c(Z-1)/A^{1/3}]/T.
  \end{eqnarray}

\section*{II.B SMM}
  In SMM, the fragmenting system is in the thermal and chemical equilibrium at low density~\cite{Bondorf95,Botvina2001,Soulioutis2007}. A Markov chain~\cite{Botvina2001} is generated to represent the whole partition ensemble in the version discussed below. All breakup channels (partitions) for nucleons and excited fragments are considered under the conservation of mass, charge, momentum, and energy. The primary fragments are described by liquid-drops at a given freezeout volume.
  Light clusters with mass number A $\le$ 4 are considered as stable particles (``nuclear gas''). Their masses and spins are taken from the experimental values. Only translational degrees of freedom of these particles are taken into account in the entropy of the system. Fragments with A $>$ 4 are treated as spherical excited nuclear liquid drops and the free energies $F_{A,Z}$ are given as a sum of the bulk, surface, Coulomb, and symmetry-energy contributions,
  \begin{eqnarray}\label{eq:FreeEnergy}
  F_{A,Z} = F^{B}_{A,Z}+F^{S}_{A,Z}+E^{C}_{A,Z}+E^{sym}_{A,Z},
  \end{eqnarray}
  where
  \begin{eqnarray}
  \label{eq:VolumeFreeE}
  F^{B}_{A,Z} &=& (-W_{0}-T^{2}/\varepsilon_{0})A,\\
  F^{S}_{A,Z} &=& B_{0}A^{2/3}\left[ \frac{T^{2}_{c}-T^{2}}{T^{2}_{c}+T^{2}} \right]^{5/4},\\
  E^{C}_{A,Z} &=& \frac{3}{5}\frac{e^2}{r_{0}}[1-(\rho/\rho_0)^{1/3}]\frac{Z^{2}}{A^{1/3}}= a^\prime_c\frac{Z^{2}}{A^{1/3}},\\
  E^{sym}_{A,Z} &=& \gamma (A-2Z)^{2}/A.
  \label{eq:SymmetryEnergy}
  \end{eqnarray}
  $W_{0}$ = 16 MeV is used for the binding energy of infinite nuclear matter, and $\varepsilon_{0}$ = 16 MeV is related to the level density; $B_{0}$ = 18 MeV is used for the surface coefficient. $T_{c}$ = 18 MeV is used for the critical temperature of infinite nuclear matter; $e$ is the charge unit and $r_0$ = 1.17 fm; $\gamma$ is the symmetry energy parameter.

  The entropy of fragments $S_{A,Z}$ can be derived from the free energy as
  \begin{eqnarray}
  S_{A,Z} = -\frac{\partial F_{A,Z}}{\partial T} = S^{B}_{A,Z}+S^{S}_{A,Z}.
  \label{eq:S_AZ}
  \end{eqnarray}
  Note that there is no symmetry entropy in Eq.~\eqref{eq:S_AZ}. According to the definition in Ref.~\cite{Natowitz10}, as shown in the appendix, the symmetry entropy depends on the density and temperature for a Fermi gas and it becomes zero for the symmetric nuclear matter. In the following, the symmetry entropy, $S_{sym}$, from Eq.~\eqref{eq:FGSsym} in the appendix at normal density is used. The symmetry free energy becomes
  \begin{eqnarray}
  F^{sym}_{A,Z} = E^{sym}_{A,Z} - TS_{sym}.
  \label{eq:Fsym}
  \end{eqnarray}

  In the micro-canonical approximation, the equation of equilibrium temperature ($T_f$) characterizing a partition $f$ is given in constraining the average energy associated with the partition by
  	\begin{eqnarray}\label{eq:temperature}
  	\forall f: \ E_f (T_f,V) = E_0,
  	\end{eqnarray}
  where $V$ and $E_0$ are the breakup volume and total energy of system, respectively. The statistical weight of the partition $f$ is calculated as
  \begin{eqnarray}\label{eq:Weight}
    W_{f} &=& \frac{1}{\xi}\exp\left[ S_f(E_0,V,A_0,Z_0) \right],
  \end{eqnarray}
  where
  \begin{eqnarray}\label{eq:Normalization}
    \xi &=& \sum_{\{f\}}\exp\left[ S_f(E_0,V,A_0,Z_0) \right],
  \end{eqnarray}
  $S_{f}$ is the entropy of the system of partition $f$, which is a function of the total energy $E_0$, mass number $A_0$, charge $Z_0$, and other parameters of the source and calculated as
  \begin{eqnarray}\label{eq:Entropy}
  \nonumber S_f &=& \sum_{A,Z}{N_{A,Z}S_{A,Z}}+S^{T}\\
  &=& \sum_{A,Z}{N_{A,Z}(S^{B}_{A,Z}+S^{S}_{A,Z}+S_{sym})}+S^{T},
  \end{eqnarray}
  where $S^T$ is the translational entropy of system and calculated as
  \begin{eqnarray}\label{eq:translationalEntropy}
  \nonumber S^{T} &=& \sum_{A,Z}{\left[N_{A,Z}\ln\left(g_{A,Z}\frac{V_f}{\lambda^3_T}A^{3/2}\right)-\ln(N_{A,Z}!)\right]}\\
   &&- \ln\left(\frac{V_f}{\lambda^3_T}A_0^{3/2}\right),
  \end{eqnarray}
  $N_{A,Z}$ is the number of fragments with mass $A$ and charge $Z$ in partition $f$, $g_{A,Z}$ is the degeneracy factor of the fragment, $\lambda_{T}$ is the nucleon thermal wavelength, $V_{f}$ is the ``free'' volume. The symmetry entropy $S_{sym}$ is added through $S_{A,Z}$ in Eq.~\eqref{eq:Entropy}.
  Since most fragments generated in the SMM simulations in this work are nearly symmetric ($(N-Z)/A \le 0.2)$, the effect of the added symmetry entropy is rather small as shown in the next section.
  No afterburner has been applied for the SMM generated events throughout the paper, thus all IMFs from the SMM calculations are the primary hot fragments.

 \section*{III. Consistency between SMM and MFM}

  In order to examine the consistency between SMM and the IYR method based on the MFM formulation described in Section II, SMM input values and those extracted by the IYR method from the SMM fragments are compared. The SMM input parameters are chosen as follows:
  the source mass number $A_{s} = 100$, charge number $Z_{s} = 45$, the fragmenting volume $V = 6V_0$.
  Source excitation energy is 7 MeV/nucleon.
  The input symmetry energy coefficient $\gamma$ varies between 0 MeV and the default value at normal density, 25 MeV.
  1 Million events are generated for each input $\gamma$.

\begin{figure}[hbt]
    \centering
    \includegraphics[scale=0.35]{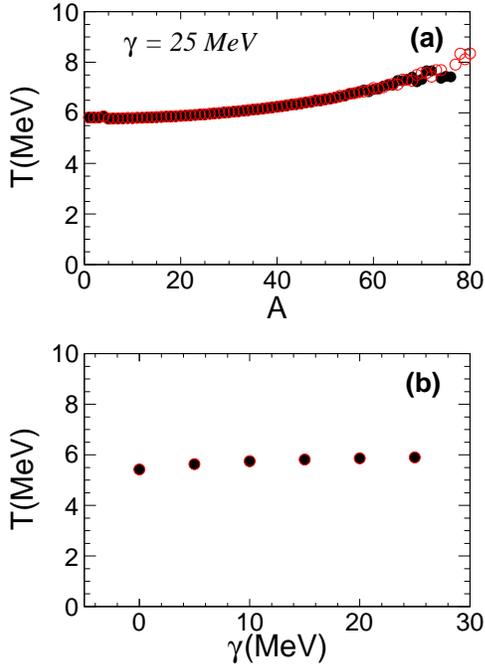}
    \caption{\footnotesize
    (Color online)(a) Quasi-temperature of SMM without (solid circles) and with (open circles) the symmetry entropy as a function of fragment mass A for $\gamma$ = 25 MeV. (b) The average temperature of SMM without (solid circles) and with (open circles) the symmetry entropy as a function of $\gamma$.
    }		
    \label{fig:fig01_Temperature}
  \end{figure}

  In SMM the ``temperature'' depends slightly on the fragmenting channel, because the energy fluctuates from partition to partition with the Markov-chain method and they are determined from the energy balance in Eq.~\eqref{eq:temperature} for a given partition. The fragment mass dependence of the quasi-temperature is shown in Fig.~\ref{fig:fig01_Temperature} (a) for $\gamma$ = 25 MeV. In Fig.~\ref{fig:fig01_Temperature} (b), the average quasi-temperature of SMM as a function of input $\gamma$ is shown. No notable changes are observed with and without the symmetry entropy.

\begin{figure}[hbt]
    \centering
    \includegraphics[scale=0.35]{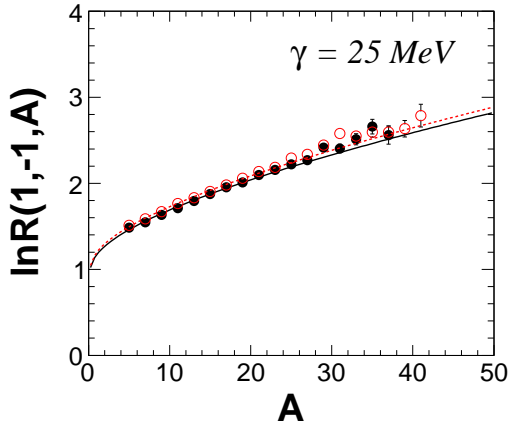}
    \caption{\footnotesize
    (Color online) $\ln[R(1,-1,A)]$ are plotted as a function of the fragment mass number A produced in SMM without (solid circles) and with (open circles) the symmetry entropy for $\gamma$ = 25 MeV. The solid and dashed lines are the fitting results with Eq.~\eqref{eq:ac_T} for SMM without and with the symmetry entropy, respectively.
    }		
    \label{fig:fig02_acT_Csym25}
  \end{figure}

  The $a_c$ value also can be extracted from the fragments generated by SMM, using Eq.~\eqref{eq:ac_T}. In Fig.~\ref{fig:fig02_acT_Csym25}, $\ln[R(1,-1,A)]$ are plotted from the fragments generated by SMM in the case of $\gamma$ = 25 MeV. Using $a_c$ and $\Delta \mu$ as free parameters in Eq.~\eqref{eq:ac_T}, the $a_c$ and $\Delta \mu$ values are extracted and the fitting results are shown by lines in the figure.

\begin{figure}[hbt]
    \centering
    \includegraphics[scale=0.35]{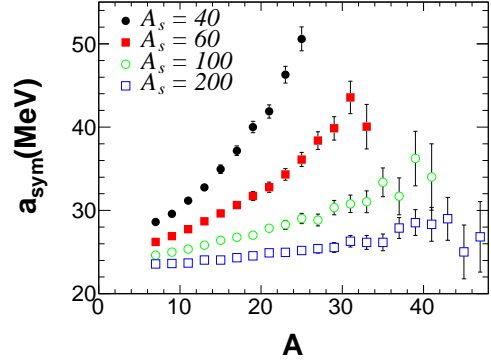}
    \caption{\footnotesize
    (Color online) Extracted $a_{sym}$ values as a function of fragment mass A for system size $A_s$ = 40 (solid circles), $A_s$ = 60 (solid squares), $A_s$ = 100 (open circles) and $A_s$ = 200 (open squares).
    }		
    \label{fig:fig03_METR}
  \end{figure}

  Using the $a_c$ value above and the quasi-temperature in Fig.~\ref{fig:fig01_Temperature}, the average $a_{sym}$ values are calculated using Eq.~\eqref{eq:asym_T} from the SMM fragments and plotted in Fig.~\ref{fig:fig03_METR} as a function of the fragment mass by green circles for the system size = 100 and $\gamma= 25$ MeV. They show a rather strong mass dependence and increase with mass. As discussed below, the mass dependence depends significantly on the system size.

\begin{figure}[hbt]
    \centering
    \includegraphics[scale=0.35]{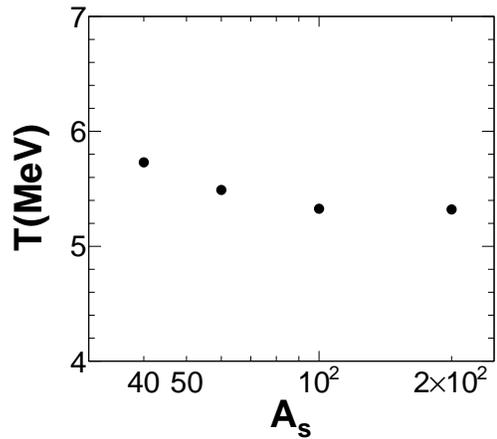}
    \caption{\footnotesize
    (Color online) Quasi-temperature of SMM with the symmetry entropy for $A_{s}$ = 40, 60, 100 and 200 with the fixed $Z_{s}/A_{s}$ = 0.45.
    }		
    \label{fig:fig04_TSMM_A40to200}
  \end{figure}

  In Ref.~\cite{Liu14PRC90_014605}, a mass dependence is observed in the extracted $a_{sym}/T$ from the experimentally reconstructed isotopes. In that analysis, the mass dependence of the $a_{sym}/T$ is attributed to the temperature, which originates from the momentum conservation during the fragmentation process.
  Since T in SMM shows almost mass independent for $A<40$ for the system size = 100 case, see Fig.~\ref{fig:fig01_Temperature} , the mass dependence of $a_{sym}$ in Fig.~\ref{fig:fig03_METR} comes from $a_{sym}$ itself. In Fig.~\ref{fig:fig03_METR}, the mass dependence of $a_{sym}$ is compared among different system sizes, $A_s$ = 40, 60, 100 and 200, with the fixed $Z_s/A_s$ = 0.45. When the system becomes larger, the mass dependence of $a_{sym}$ becomes less and it becomes closer to the input $\gamma$ value of 25 MeV. In Fig.~\ref{fig:fig04_TSMM_A40to200}, the average temperature values are plotted as a function of the system mass $A_s$ with the fixed $Z_s/A_s=0.45$. The results show that the temperature has a system size dependence in SMM. The decreasing trend as increasing the system size reflects the fact that the mass dependence becomes less for the heavier fragmenting system.

  From these facts, we concluded that the mass dependence observed in the extracted $a_{sym}$ originates from a system size effect and we call it ``finite size effect". The correction made for the effect is called ``finite size correction'' throughout the paper.

  \begin{figure}[hbt]
    \centering
    \includegraphics[scale=0.35]{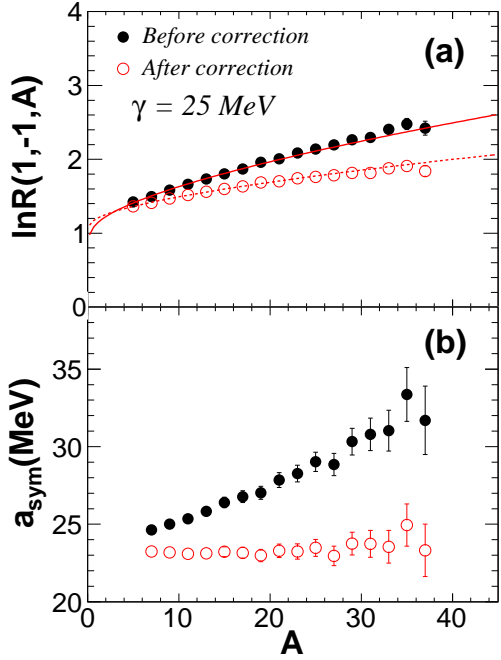}
    \caption{\footnotesize
    (Color online) (a) $\ln[R(1,-1,A)]$ without (solid circles) and with (open circles) the finite size correction are plotted as a function of the fragment mass number A for SMM with the symmetry entropy. The solid and dashed lines are the fitting results of Eq.~\eqref{eq:ac_T}. (b) Extracted average $a_{sym}$ values without (solid circles) and with (open circles) the finite size correction as a function of the fragment mass number A for SMM with the symmetry entropy.
    }		
    \label{fig:fig05_FiniteCorrection}
  \end{figure}

  \begin{figure}[hbt]
    \centering
    \includegraphics[scale=0.45]{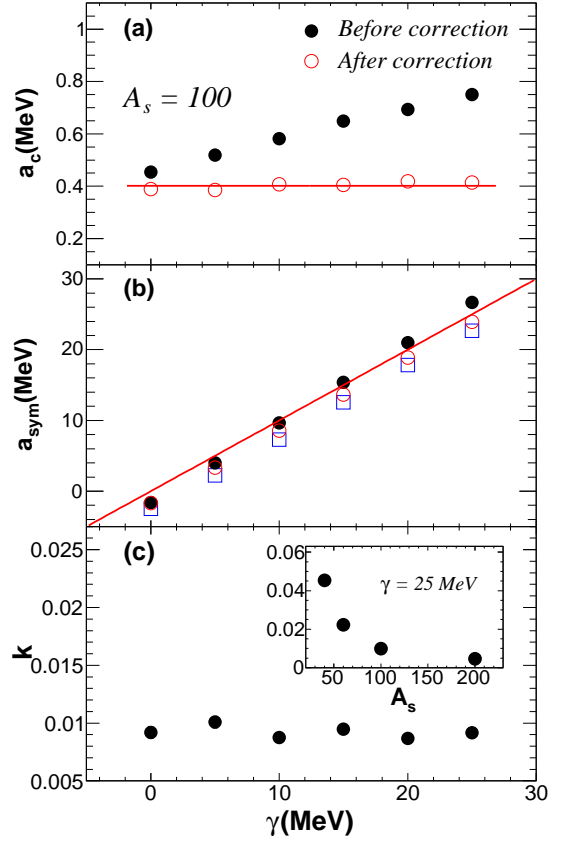}
    \caption{\footnotesize
    (Color online) (a) Extracted $a_c$ values as a function of input $\gamma$ before (solid circles) and after (open circles) the finite size correction for SMM with symmetry entropy. The fitting errors are smaller than the symbol size. The line is from a constant fit. (b) Average $a_{sym}$ values of the constant fit as a function of input $\gamma$ before (solid circles) and after (open circles) the finite size correction for SMM with symmetry entropy. The open squares are those after the finite size correction for SMM without the symmetry entropy. The line is corresponding to the SMM input value of $a_{sym} = \gamma$.
    (c) Extracted $k$ values are plotted as a function of $\gamma$ in the case of system size = 100. The $k$ value for $\gamma$ = 0 MeV is averaged over those of $\gamma >$ 0 MeV. The inset shows the extracted $k$ values as a function of the system size for the case of $\gamma$ = 25 MeV.
    }		
    \label{fig:fig06_FiniteCorrectionSummary}
  \end{figure}

  In order to take into account the finite size effect for the SMM events, the free energy in MFM is modified as
  \begin{eqnarray}\label{eq:FiniteSizeFactor}
  \nonumber Y(A,Z) &=& Y_{0}A^{-\tau}\\
  \nonumber & &\exp \left\{\left[ \frac{W(I,A) + \mu_n N + \mu_p Z}{T_{SMM}}
  +S^{mix}_{A,Z} \right]\right.\\
  & &\left. (1+kA)\right\},
  \end{eqnarray}
  where $T_{SMM}$ is the quasi-temperature from SMM. The $k$ value is optimized to make the $a_{sym}$ mass independent for a given system size and a given $\gamma$ value. The finite size correction is made for events generated at all $\gamma$ values except for $\gamma$ = 0 MeV. For the case of $\gamma$ = 0 MeV, no mass dependence is observed for the extracted $a_{sym}$ values.
  One should note that the correction made in Eq.~\eqref{eq:FiniteSizeFactor} is not universal and should only apply for the SMM generated events.
  Fig.~\ref{fig:fig05_FiniteCorrection} shows the results for $\ln[R(1,-1,A)]$ in (a) and the extracted $a_{sym}$ in (b) as a function of the fragment mass A. Solid and open circles show the results before and after the correction in the case of $\gamma$ = 25 MeV, respectively.
  For different $\gamma$ values, the extracted $a_c$ before (solid circles) and after (open circles) the correction are plotted in Fig.~\ref{fig:fig06_FiniteCorrectionSummary} (a) for SMM with symmetry entropy in the case of $A_s$ = 100 and $Z_s/A_s$ = 0.45. The extracted $a_c$ values are almost constant for different $\gamma$ values after the finite size correction and
  the average $a_c$ value with standard deviation is $\langle a_c\rangle = 0.40 \pm 0.01$ MeV.  This value is slightly larger than the SMM input Coulomb energy coefficient under the Wigner-Seitz approximation of  $a^\prime_c = 1.44\times3/(5r_{0})[1-(V_0/V)^{1/3}] = 0.33$ MeV for $V=6V_0$. This issue is further discussed in Section IV.
  In Fig.~\ref{fig:fig06_FiniteCorrectionSummary} (b) the extracted $a_{sym}$ are compared with the input values (line).
  All extracted average $a_{sym}$ before and after the correction are distributed around the input $\gamma$ values (line) for different $\gamma$ values and agree with the input values within $\sim$ 2 MeV.
  The $k$ value is determined for each $\gamma$ value except for $\gamma$ = 0 MeV, which is the value averaged over those of $\gamma > 0$ MeV. The extracted $k$ values are plotted as a function of $\gamma$ in Fig.~\ref{fig:fig06_FiniteCorrectionSummary} (c). As shown in the inset, the $k$ value decreases significantly as the system size, $A_{s}$, increases. This reflects the fact that the system mass dependence of $a_{sym}$ values becomes smaller when $A_{s}$ becomes larger as shown in Fig.~\ref{fig:fig03_METR}. The $k$ values are distributed around a constant value of 0.01 for different $\gamma$ values as shown in Fig.~\ref{fig:fig06_FiniteCorrectionSummary} (c) in the case of $A_s$ = 100.

\section*{IV. Statistical analysis of the reconstructed experimental data}

  In this section, the experimental data from the reconstructed isotopes are compared with SMM simulated events. In our previous works~\cite{Lin14,Lin14PRC89_021601R}, the primary isotope yields were experimentally reconstructed in the $^{64}$Zn + $^{112}$Sn reaction at 40 MeV/nucleon. These yields allow us to compare directly to the SMM primary fragments without an afterburner. The SMM calculations are performed with source size $A_{s}$ = 60, charge number $Z_s$ = 27, which are extracted from the NN source component of the experimentally observed energy spectra for all particles, including neutrons~\cite{Wada13}. The source excitation energy is calculated using the temperature from self-consistent analysis~\cite{Lin14,Lin14PRC89_021601R} and the fragment multiplicities $M_i$ as

  \begin{equation}
  E^{*} = \sum_{i} (3/2)T M_i  - Q.
  \label{eq:Ex_hot}
  \end{equation}

  For the multiplicity, the experimental values of the NN source from the cold light particles (LPs) and fragments are used, since the reconstructed primary LP multiplicity values are not available. LPs' contribution dominates in Eq.~\eqref{eq:Ex_hot}. Q is an average Q value. $E^{*}$ = 6.7 MeV/nucleon is obtained from the experimentally extracted temperature value of T = 5.9 MeV~\cite{Lin14PRC89_021601R,Lin14}. $\gamma$ = 20.7 MeV from the self-consistent analysis is used. In Refs.~\cite{Lin14,Lin14PRC89_021601R}, the density of the fragments at the time of the fragments formation $\rho/\rho_0$ = 0.54 is extracted. However this density is the average density inside fragments and different from the SMM density, which represents the density for the whole system at the time of the fragmentation. In SMM, no solution was found for the fragment partition at $V/V_0 \le 3$. Therefore in simulations below, the breakup volume of $V/V_0$ = 4, 6, 10 are examined.

  \begin{figure}[hbt]
	\centering
	\includegraphics[scale=0.35]{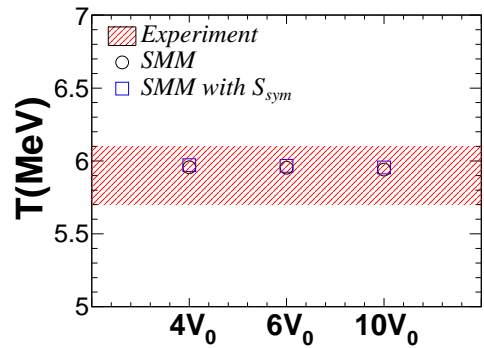}
	\caption{\footnotesize
		(Color online) Average temperature from self-consistent analysis (shaded area) from Refs.~\cite{Lin14,Lin14PRC89_021601R} and that from SMM with $V/V_0$ = 4, 6 and 10. Open circles and open squares are without and with the symmetry entropy, respectively.}		
	\label{fig:fig07_Temperature_Zn64Sn112}
  \end{figure}

  \begin{figure}[hbt]
    \centering
    \includegraphics[scale=0.4]{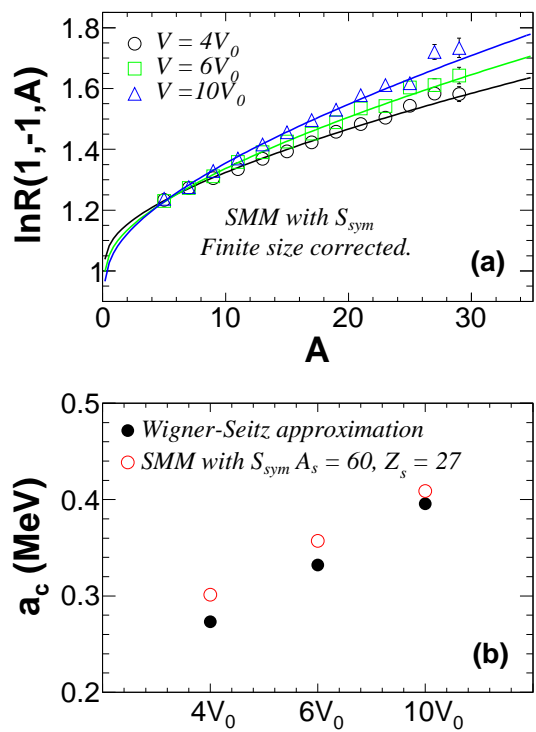}
    \caption{\footnotesize
    (Color online)
    (a) $\ln[R(1,-1,A)]$ as a function of fragment mass from IMFs generated by SMM for different breakup volumes of $V/V_0$ = 4, 6 and 10. Lines are the fitting results using Eq.~\eqref{eq:ac_T} for each breakup volume. (b) Open circles are the extracted $a_c$ values from the fitting in (a) and solid circles are the SMM input values under the Wigner-Seitz approximation.}
    \label{fig:fig08_Coulomb_Zn64Sn112}
  \end{figure}

  In Fig.~\ref{fig:fig07_Temperature_Zn64Sn112}, the temperature from the self-consistent analysis (shaded area) and that from SMM at different breakup volumes (symbols) are compared. The SMM temperature values are nearly constant for different breakup volumes and agree well with the temperature from the self-consistent analysis of the reconstructed isotope yields.
  The Coulomb energies extracted in the same way as those in the previous section are also compared.
  In Fig.~\ref{fig:fig08_Coulomb_Zn64Sn112} (a), $\ln[R(1,-1,A)]$ are plotted from IMFs in the SMM events for different breakup volumes. The finite size effect has been taken into account. For each breakup volume, $a_c$ and $\Delta_{\mu}$ values are extracted as free parameters, using Eq.~\eqref{eq:ac_T}. The $a_c$ value relates to the different curvatures in Fig.~\ref{fig:fig08_Coulomb_Zn64Sn112} (a) and one can see clear differences in the figure. The extracted $a_c$ values are plotted in (b) for different breakup volumes (open circles). The solid circles represent the SMM input values which are calculated under the Wigner-Seitz approximation. The extracted $a_c$ values are $\sim 0.02-0.03$ MeV larger than those of the input values, but the increasing trend as a function of different breakup volumes is well reproduced.

  \begin{figure}[hbt]
  	\centering
  	\includegraphics[scale=0.45]{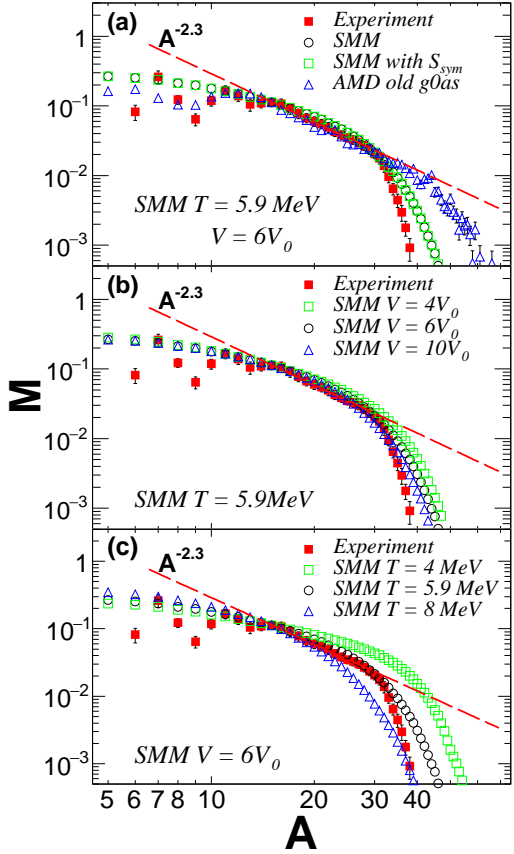}
  	\caption{\footnotesize
  		(Color online) (a) The experimental mass distribution (solid squares) is compared with that of SMM without (open circles) and with (open squares) the symmetry entropy at $T=5.9$ MeV and the breakup volume of $6V_0$. The mass distribution of AMD from Refs.~\cite{Lin14,Lin14PRC89_021601R} is also shown by triangles. The distributions of the simulated results are normalized to the reconstructed data at A = 15. (b) The experimental mass distribution is compared with that of SMM with different breakup volumes at $T=5.9$ MeV. (c) The experimental mass distribution is compared with that of SMM with different temperatures at $V=6V_0$.
  	}		
  	\label{fig:fig09_Mass_Zn64Sn112}
  \end{figure}

  In Fig.~\ref{fig:fig09_Mass_Zn64Sn112} (a), mass distribution of the experimentally reconstructed isotopes (solid squares) is compared with the simulations. The results for SMM with and without the symmetry entropy are almost identical (open squares and circles). AMD results from Refs.~\cite{Lin14,Lin14PRC89_021601R} are also plotted (open triangles). All calculated yields are normalized to that of the reconstructed data at A = 15. They reproduce the experimental primary mass distribution for fragments with $ 10 < A < 30$ reasonably well.
  The experimental yields in $A < 10$ show a significant uneven structure, but all calculated results show rather smooth distributions. The experimental uneven structure is partially caused by unstable nuclei, such as $^{8}Be$ and $^{9}B$, which are included in the calculated yields, but for those such as $A=6$ or $10$ the reason is unknown. The experimentally observed power law distribution of fragment yields with the exponent of $\tau = -2.3$ in $ 10 < A < 30$ is also held for all simulations. The deviation from the power law line for the reconstructed data above $A = 30$ is partially caused by the experimental limitation of the available identified isotopes, which were used for the reconstruction ($Z \le 15$). The simulated fragment yields do not have such limitations, but SMM results show a similar trend as that of the experiment. The deviation of the AMD results is much less as the mass increases.
  We also investigate the effects of breakup volume and temperature in the SMM. The experimental mass distribution is compared with those from the SMM events at different breakup volumes in Fig.~\ref{fig:fig09_Mass_Zn64Sn112} (b) and at different breakup temperatures in Fig.~\ref{fig:fig09_Mass_Zn64Sn112} (c). The SMM mass distribution is not sensitive to the breakup density. On the contrary it is very sensitive to the breakup temperature. The best result is obtained at $T \sim$ 6 MeV which is consistent to the experimentally determined temperature value of $T$ = 5.9 MeV in Refs.~\cite{Lin14,Lin14PRC89_021601R}.

  \begin{figure}[hbt]
    \centering
    \includegraphics[scale=0.43]{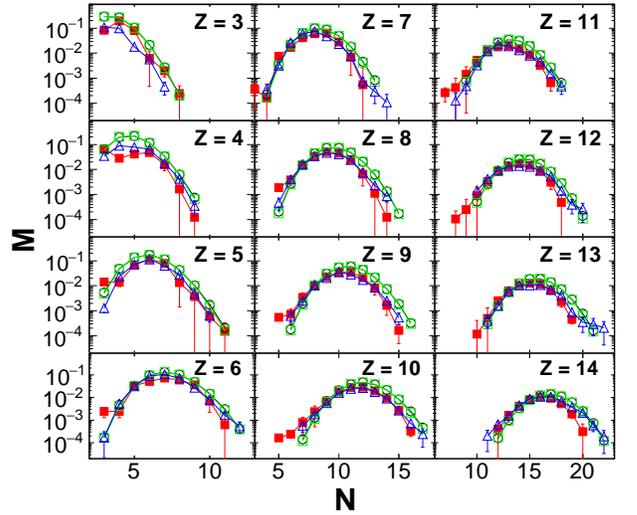}
    \caption{\footnotesize
    (Color online) Isotope distributions of the experimentally reconstructed primary fragments (solid squares) and those from SMM without (open circles) and with (open squares) the symmetry entropy at $V=6V_0$ are compared for $Z = 3 - 14$. AMD results from Refs.~\cite{Lin14,Lin14PRC89_021601R} are also shown by open triangles. All results are plotted in an absolute scale.
    }		
    \label{fig:fig10_Isotopes_Zn64Sn112}
  \end{figure}

   In Fig.~\ref{fig:fig10_Isotopes_Zn64Sn112} detail comparison of isotope yield distributions are carried out in an absolute scale for $Z = 3 - 14$ between the experimentally reconstructed primary isotopes and the fragments from the SMM events at $V=6V_0$ without (open circles) and with (open squares) the symmetry entropy. AMD results from Ref.~\cite{Lin14} are also shown by open triangles.
   Reasonable agreements are found between the SMM calculations and the reconstructed data, but the widths of the SMM distributions are slightly wider than the experimental ones for all Z values, whereas those of AMD simulations reproduce the widths slightly closer to those of the experimental distributions.  The significant differences in the simulated results for Z = 4 are caused by the fact that $^{8}$Be was missing among the final secondary products in the reconstruction, which is crucial for Z = 4 primary fragments.

\begin{figure}[hbt]
    \centering
    \includegraphics[scale=0.35]{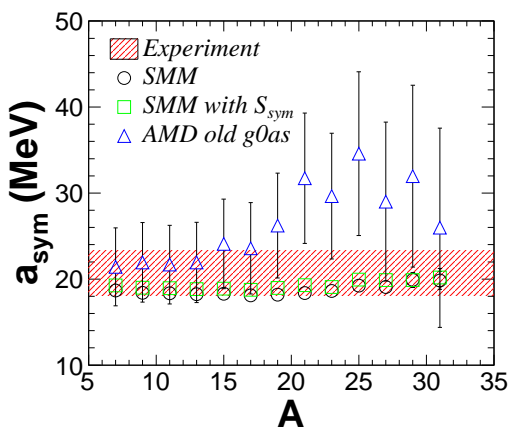}
    \caption{\footnotesize
    (Color online) $a_{sym}$ as a function of fragments mass A for the reconstructed data (shaded area) and the SMM results without (open circles) and with (open squares) the symmetry entropy at $V = 6V_0$ are shown, together with those of AMD from Refs.~\cite{Lin14,Lin14PRC89_021601R}.
    }		
    \label{fig:fig11_asym_Zn64Sn112}
  \end{figure}

   In order to see the consistency of symmetry energy coefficient between the reconstructed data and the simulation events, we apply Eq.~\eqref{eq:asym_T} both to the SMM and experimental isotope yields. Fig.~\ref{fig:fig11_asym_Zn64Sn112} shows the extracted $a_{sym}$ from the SMM fragments and those of the experiment as a function of the fragment mass. The extracted values are consistent to those extracted from the reconstructed data within the error bars shown by the shaded area. AMD results from Ref.~\cite{Lin14} are also plotted by open triangles. The larger errors for the AMD results are because of the poor statistics.


   The reasonable agreements between the results from the reconstructed experimental data and those from the SMM fragments, shown in Figs.~\ref{fig:fig09_Mass_Zn64Sn112} to \ref{fig:fig11_asym_Zn64Sn112}, strongly suggest that the experimentally observed IMFs originate from a thermally and chemically equilibrated source at a freezeout volume through a multifragmentation process. This is consistent to our previous results obtained in Ref.~\cite{Liu15}, in which the existence of the freezeout volume for the IMF production is suggested from the AMD simulated events from 35 to 300 MeV/nucleon, using the IYR technique and the self-consistent method.

\section*{V. Summary }

  Firstly the consistency between SMM and MFM is examined, using the IYR technique based on MFM. The extracted $a_c$ and $a_{sym}$ values from the SMM fragments are consistent to the SMM input values after the system size effect is taken into account, though tiny deviations are also observed. The newly added symmetry entropy does not affect the results very much, because most isotopes generated in this work has $(N-Z)/A \le 0.2$ and the symmetry entropy is close to zero. Utilizing the experimentally extracted temperature and symmetry energy, SMM is applied to $^{64}$Zn + $^{112}$Sn reaction at 40 MeV/nucleon. Experimentally observed primary fragment mass and isotope distributions are compared with those of SMM at three breakup volumes (4$V_0$, 6$V_0$ and 10$V_0$). Good agreements are observed at $T \sim 6$ MeV and $\gamma \sim 20$ MeV for all break up volumes. The extracted $a_{sym}$, using the IYR technique both from these SMM events and the reconstructed IMFs from experimental data are also consistent. These agreements strongly suggest that the experimentally observed IMFs originate from a thermally and chemically equilibrated source at a freezeout volume through a multifragmentation process.

\section*{Acknowledgments}

  The authors thank A. S. Botvina for providing his code and many fruitful discussions. This work is supported by the National Natural Science Foundation of China (Grant No. 91426301 and No. 11075189), the Strategic Priority Research Program of the Chinese Academy of Sciences ``ADS project'' (Grant No. XDA03030200) and the Program for the CAS ``Light of West China'' (No. 29Y601030). This work is also supported by the US Department of Energy under Grant No. DE--FG02--93ER40773. One of the author (R.W) thanks the program of the ``visiting professorship of senior international scientists of the Chinese Academy of Sciences'' for their support during his stay in IMP.

\section*{Appendix. Symmetry entropy for a Fermi gas system}
  For an ideal Fermi gas, the average number of fermions in a single-particle state $i$ is given by the Fermi-Dirac distribution as
  \begin{equation}
  f_{i} = \frac{1}{e^{(\epsilon_i-\mu)/T}+1},
  \label{eq:FD_distribution}
  \end{equation}
  where $T$ is the temperature, $\epsilon_i$ is the energy of the single-particle state $i$, and $\mu$ is the chemical potential. The number of states between $\epsilon$ and $\epsilon+d\epsilon$ is
  \begin{equation}
  D(\epsilon)d\epsilon = g\frac{2\pi V}{h^3}(2m_0)^{3/2} \epsilon^{1/2}d\epsilon,
  \label{eq:NoStates}
  \end{equation}
  where $g$ is the degeneracy factor, $V$ is the system volume and $m_0$ is the mass of the fermion. The density $\rho$, total number A and energy U of the free Fermi gas are given by
  \begin{eqnarray}
  \label{eq:FGdensity}
   \rho &=& g\frac{2\pi}{h^3}(2m_0T)^{3/2}\int^\infty_0\frac{x^{1/2}dx}{e^{x-\mu/T}+1},\\
      A &=& g\frac{2\pi V}{h^3}(2m_0T)^{3/2}\int^\infty_0\frac{x^{1/2}dx}{e^{x-\mu/T}+1},\\
      U &=& g\frac{2\pi V}{h^3}(2m_0T)^{3/2}T\int^\infty_0\frac{x^{3/2}dx}{e^{x-\mu/T}+1}.
  \end{eqnarray}
  Then the entropy of the free Fermi gas is given as
  \begin{eqnarray}
  \nonumber \frac{S(A)}{A} &=& \frac{U-F}{AT}\\
  \nonumber  &=& \frac{U+PV - \mu A}{AT}\\
  \nonumber  &=& \frac{\frac{5}{3}U-\mu A}{AT}\\
    &=& \frac{5}{3}\frac{\int^\infty_0\frac{x^{3/2}dx}{e^{x-\mu/T}+1}}{\int^\infty_0\frac{x^{1/2}dx}{e^{x-\mu/T}+1}}- \frac{\mu}{T},
  \label{eq:FGSsystem}
  \end{eqnarray}
  where $F = \mu A - PV$ is the free energy of system and $P = \frac{2}{3}\frac{\partial U}{\partial V}$ is the pressure of fermion system.

\begin{figure}[hbt]
    \centering
    \includegraphics[scale=0.38]{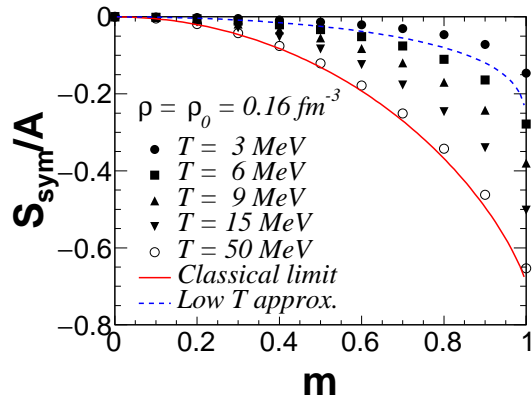}
    \caption{\footnotesize
    (Color online) Symmetry entropy per nucleon, $S_{sym}/A$, as a function of $m$ for $T$ = 3 MeV (solid circles), 6 MeV (solid squares), 9 MeV (solid up triangles), 15 MeV (solid down triangles) and 50 MeV (open cirlces) at density $\rho = \rho_0$. Solid line corresponds to the symmetry entropy at the classical limit, and dashed line represents the results using Eq.~\eqref{eq:SsymLowT}.
    }		
    \label{fig:fig12_Ssym}
  \end{figure}

   According to Ref.~\cite{Natowitz10}, the symmetry entropy is defined as the difference between the entropies of pure proton or neutron and symmetric nuclear matter. For a nuclear system with A nucleons (N neutrons and Z protons), therefore, the symmetry entropy per nucleon is calculated as
  \begin{eqnarray}
  \frac{S_{sym}}{A} = \frac{S_{A,Z}^{B,tot}}{A} - \frac{S_{A,A/2}^{B,tot}}{A},
  \label{eq:FGSsym}
  \end{eqnarray}
  where $\frac{S_{A,Z}^{B,tot}}{A} = \frac{1+m}{2}\frac{S(N)}{N} + \frac{1-m}{2}\frac{S(Z)}{Z}$ is the average entropy of $N$ neutrons and $Z$ protons system. $m=\frac{N-Z}{A} = \frac{\rho_n-\rho_p}{\rho}$ is the asymmetry parameter. One should note that $S_{A,A/2}^{B,tot}$ is the volume entropy taken into account in the second term in the right hand side of Eq.~\eqref{eq:VolumeFreeE}, $S_{A,Z}^B=S_{A,A/2}^{B,tot}$. The calculated symmetry entropy per nucleon as a function of $m$ is shown in Fig.~\ref{fig:fig12_Ssym} at density $\rho=\rho_{0}$ and different temperatures. The solid line represents the symmetry entropy per nucleon at the classical limit, which is given analytically as
  \begin{eqnarray}
   \frac {S_{sym}}{A}
    = - \left[\frac{N}{A}\ln(N/A)+\frac{Z}{A}\ln(Z/A)\right] - \ln(2).
  \label{eq:SsymCL}
  \end{eqnarray}

  In Eq.~\eqref{eq:FGSsym} the exact derivation of the symmetry entropy from a Fermi gas is used. However in SMM, the bulk entropy in Eq.~\eqref{eq:VolumeFreeE} is derived, using the low temperature approximation. In order to verify the consistency in the above discussion with the exact quantum symmetry entropy, the approximated symmetry entropy with the same low temperature approximation used in SMM, which reads as
  \begin{eqnarray}
  \frac {S_{sym}}{A} = \frac{\pi^2}{2}\frac{T}{\varepsilon_F}\left[\frac{1}{2}(1+m)^{1/3}+\frac{1}{2}(1-m)^{1/3}-1\right],
  \label{eq:SsymLowT}
  \end{eqnarray}
  where $\varepsilon_F =$ 36.8 MeV at $\rho = \rho_0$, is also shown in dashed line in Fig.~\ref{fig:fig12_Ssym} at $T$ = 6 MeV. The approximated symmetry entropy shows slightly higher value than the exact quantum one.

\end{document}